# Credibility in Web Search Engines

**Dirk Lewandowski**

*Hamburg University of Applied Sciences, Germany*

## ABSTRACT

Web search engines apply a variety of ranking signals to achieve user satisfaction, i.e., results pages that
provide the best-possible results to the user. While these ranking signals implicitly consider credibility
(e.g., by measuring popularity), explicit measures of credibility are not applied. In this chapter, credibility
in Web search engines is discussed in a broad context: credibility as a measure for including documents in
a search engine's index, credibility as a ranking signal, credibility in the context of universal search
results, and the possibility of using credibility as an explicit measure for ranking purposes. It is found that
while search engines—at least to a certain extent—show credible results to their users, there is no fully
integrated credibility framework for Web search engines.

## INTRODUCTION

Search engines are used for a wide variety of research purposes, and they are often the first place to go
when searching for information. Users select search results and then read Web pages based on their
decisions about whether the information presented by the search engine is of value to them. Search
engines are so popular that they are, together with e-mail, the most commonly used service on the Internet
(Purcell, 2011). Every day, billions of queries are entered into the search boxes of popular search engines
such as Google and Bing. Therefore, these engines play an important role in knowledge acquisition, not
only from an individual's point of view, but also from society's point of view. It is astonishing to what
degree users trust search engines (e.g., (Hargittai, Fullerton, Menchen-Trevino, & Thomas, 2010; Pan et
al., 2007), and they rely on Web search engines to display the most credible results first.

Search engine rankings, however, do not guarantee that credible pages are ranked first for every topic.
The construct underlying the rankings of search engines basically assumes that a page's popularity equals
credibility, although other factors also play a role. Popularity, in this case, refers to popularity among all
Web authors and readers by measuring the distribution of links, clicks within the search engine results
pages, time spent reading the results documents, and recommendations in social media. While applying
these factors in ranking algorithms often leads to good results, it should be stressed that these popularity
measures always rely on the users judging the credibility of the documents, i.e., people only make
credible pages popular.

Technical means for finding suitable indicators for credible Web pages are an alternative to human
credibility judgments about Web search engine results (Mandl, 2005, 2006). Apart from popularity
analyses, page and text properties can be used to estimate the credibility of a document, although such
approaches can only provide estimates. Before discussing credibility further, we first need to define the
concept in the context of search engine results.

According to the Encyclopedia of Library and Information Sciences (Rieh, 2010), credibility is an
intuitive and complex concept that has two key dimensions: trustworthiness and expertise. Both are
judged by people consuming information, and therefore, credibility always lies in the eye of the beholder.
In our discussion of credibility in the context of Web search engines, we follow Rieh's definition:
"Credibility is defined as people's assessment of whether information is trustworthy based on their own expertise and knowledge" (p. 1338). However, as search engines rate documents algorithmically, we need to consider "people" not only being users of information, but also designers of search engines and their ranking algorithms, who have certain assumptions about credibility that are then used in the system.

To get a deeper understanding of credibility, we refer to Tseng and Fogg's (1999) four types of credibility:

1. Presumed credibility, where people have general assumptions about a source of information (e.g., assuming that a friend will tell them the truth, or that articles written by full-time journalist will give credible information)
2. Reputed credibility, where sources of information are seen as credible because third parties assigned credibility to them in the past. E.g., the title of doctor or professor makes most people believe that this person is a credible source of information.
3. Surface credibility, where credibility is given to a source of information because of surface criteria, such as the jacket design of a book or the layout of a webpage.
4. Experienced credibility, where the person judging credibility has first-hand experience with the source of information.

In information retrieval (IR) evaluations, users judge the relevance of documents linked to a search query or information need in a controlled environment. However, while the concept of relevance somehow incorporates credibility, it also incorporates many other aspects. If expert jurors who are instructed to research the credibility of the documents are asked, then statements about credibility can be made. However, such studies are rare, mainly because expert jurors are expensive and the process of evaluating credibility is time-consuming. Furthermore, there is no tradition of credibility evaluation in IR because in traditional IR systems (e.g., newspaper databases or patent collections), the quality of the documents is controlled in the process of producing the database (Rittberger & Rittberger, 1997), and only documents from collections deemed credible are included.

Information quality frameworks (e.g., Knight & Burn, 2005) are of only limited use in the context of Web search engines, as the main problem is that search engine users apply credibility judgments when considering (1) the results descriptions ("snippets") on the search engine results pages, and (2) the results documents themselves. In both cases, they have only limited resources for judging credibility. It is much easier to apply information quality criteria to the inclusion of documents into an information system than applying such criteria to the ranking of documents, or even to rely on the system's users to judge the quality of the documents.

Already from this short introduction, we can see that credibility in Web search engines has multiple dimensions; it is a concept that, while it is inherent in search engine rankings and users' perceptions of search engines, has not yet been fully explored. Therefore, the aim of this chapter is to clarify the meaning of credibility in the context of search engines and to explore where credibility is applied in this context.

This chapter is structured as follows. First, the criteria by which search engines decide upon including documents in their indices are discussed. Then, we consider ranking signals generally applied in ranking Web documents, and show that while credibility is a measure implicitly considered in these rankings, it is mainly achieved through measuring popularity. As the positions of search engine results highly influence the probability of these results being selected, it is of interest to content providers to optimize their documents for ranking in search engines. How such optimization can influence the overall credibility of results will be discussed under a separate heading. Then, we will examine results' credibility when not only considering general-purpose ranking of search results, but also the inclusion of so-called universal search results into the search engine results pages, i.e., results from specially curated collections. After that, we will turn to search engine evaluation, and how credibility criteria are and could be applied to it. Finally, we discuss the search engines' responsibility to provide credible results. The chapter concludes with some suggestions for future research.

## SEARCH ENGINES' INCLUSION CRITERIA

Search engines are usually seen as tools that aim to index "the whole of the Web". While this surely is not achievable from a technical, as well as from an economic standpoint (Lewandowski, 2005), there are also credibility reasons for not indexing every document available on the Web. While the Web offers a great variety of high-quality information, one should not forget that a large ratio of the documents offered on the Web is of low quality. While there are no exact numbers on how many pages on the Web can be outright considered as Web spam, the wide variety of spamming techniques facing search engines, as described in the literature, shows that protecting users against this type of content is a major task for search engines (Gyongyi & Garcia-Molina, 2005; Berendt, 2011). It is also understandable that search engine vendors do not provide much information on that issue, as doing so would invite spammers to flood search engine indices, and as techniques for fighting spam are also trade secrets, which are of high value to the search engine vendors.

While search engines do not explicitly judge the *contents* of documents on the Web, they do decide against the inclusion of documents from known low-quality sources. Decisions against including content in their indices are always based on the source, not on the individual document. However, even though lots of low-quality documents are included in the search engines' databases, this does not necessarily mean that a user gets to see any of them, because of elaborate ranking functions.

However low the barriers for including contents in Web search engines' databases are, they still exist. Consider, for example, pages with spyware or other types of malware. These pages are either excluded from the index entirely, or are specially marked when displayed on a search engine's results page (McGee, 2008).

This "hard" credibility measurement is *explicit* in that such documents are either not included in the index at all, or the users are warned when such a document is displayed in the results list. On the other hand, *implicit* judgments on credibility are only shown through the results ranking, which is not comprehensible to the general user.

To summarize, credibility criteria are implemented in search engines in three different ways:

1. Pages of low credibility are excluded from the search engines' indices.
2. Pages of low credibility are specially marked in the results presentation.
3. Pages of low credibility are ranked lower in the results lists.

Again, it should be stressed that while search engines do apply criteria for including documents in their indices, the barriers are very low, and mainly documents or document collections only built for the purpose of being included in the search engines' indices for supporting other, to-be optimized documents through links are excluded. While in some particular cases, a user might miss relevant documents explicitly searched for because the search engine used simply did not index them, this now occurs rather rarely.

## DOES POPULARITY EQUAL QUALITY?

It is a mistake to think that in search engines, credibility does not play a role in ranking. However, while search engines do not measure credibility explicitly, measures of credibility are surely implicit of other measures, such as popularity.

Before discussing the influence of credibility on search engine rankings in detail, we will give a short overview of the ranking factors applied in search engine rankings. Ranking algorithms can be broken down into individual signals, and they can consist of hundreds of these signals. However, these signals can be grouped into a few areas, as follows:

1. Text-based matching: Simple text-based matching, as applied in all text-based information retrieval systems, matches queries and documents to find documents that fulfill the query. Text-based ranking factors such as term frequency / inverted document frequency (TF/IDF) are based on assumptions about the occurrence of terms in a document (e.g., an ideal "keyword density") and allow for a ranking that differentiates not only between documents containing the keywords entered and ones where the keywords are not present, but also between document weights; this allows for a rank-based list of results. However, such text-based ranking algorithms are designed for collections where all documents are deemed credibly (e.g., newspaper databases where quality control is applied before individual documents are added to the database). They fail in the case of the Web, where content providers are able to manipulate documents for gaming the search engines. Therefore, while text-based matching forms the basis of search engine rankings, additional factors measuring the quality of the documents are required.

2. Popularity: The popularity of a document is referenced for its quality evaluation. For example, the number of user accesses and the dwell-time on the document is measured, as well as the linking of a document within the Web graph, which is decisive for the ranking of Web documents. For this purpose, not only are the number of clicks and links crucial, but weighted models are also implemented that enable a differentiated evaluation. These models are well documented in the literature (Culliss, 2003; Dean, Gomes, Bharat, Harik, & Henzinger, 2002; Kleinberg, 1999; Page, Brin, Motwani, & Winograd, 1998) and are still considered most important in search results rankings (Croft, Metzler, & Strohman, 2010). Popularity-based measures can be divided into three groups:

   a. Link-based measures: The classic method to determine popularity is through links. A link pointing to a Web page can be seen as a vote for that page, and when weighting links according to the authority of the linking page, good measurements can be achieved.

   b. Click-based measures: Using click-based measures to determine quality has the advantage that such measures are available almost immediately, while link-based measures require time to build up. The drawback, however, is that in search engines, most users only click on the results presented first; therefore, click-based measures are heavily biased, as not every document even has the opportunity to be selected.

   c. Social signals: In the context of social media, explicit ratings of documents are ubiquitous.[1] These judgments can be exploited for ranking, assuming that the search engine has access to data from a social network.

3. Freshness: The evaluation of freshness is important for Web search engines in two respects. Firstly, it is a matter of finding the actual or rather relative publication and refresh dates (Acharya et al., 2005). Secondly, the question concerns in which cases it is useful to display fresh documents preferentially.

4. Locality: Knowing the location of an individual user is of great use for providing relevant results. This not only holds true in a mobile context, but also for desktop use.

5. Personalization: The aim to provide users with tailored results is referred to as personalization and combines measures from the user's own behavior (through queries entered, results selected, reading time) with measures from other users' behavior (focusing on the one hand on all users, and on the other hand on the users socially connected to the user in question), and with general measures (freshness and locality).

From this short explanation of search engine ranking signals, we can conclude that popularity lies at the heart of these systems, whether such popularity exists with all the Web's content producers (who set links to other pages and thus determine their popularity), with a certain user group (e.g., the contacts of an individual user), or with an individual user (through his clicks and viewing patterns).

The question that arises from the discussion of ranking signals is how search engines are able to show credible results without explicitly considering credibility in the documents. When looking at credibility or, more generally, information quality frameworks (Knight & Burn, 2005; Wang, Xie, & Goh, 1999; Xie, Wang, & Goh, 1998), we can see that the criteria generally mentioned are not easily applicable to algorithms. Therefore, "workarounds" must be found.

---

[1] There are also implicit social signals, but as these are also click-based, they fall under c.

In Table 1, measures used to determine the credibility of documents are shown. Credibility can be derived through the analysis of the source, the users' selection behavior, recommendations through links, and explicit ratings in social media. All these can be measured in different ways, and in the cases where users are considered, they can further be differentiated according to the group of users taken into account.

| Credibility through… | Measures | Based on… |
|---|---|---|
| Source | Domain popularity | Link graph |
| Selection behavior | • Click-through rate, i.e., how often a certain document is selected when shown<br>• Time spent reading when document was selected<br>• Bounce rate, i.e., how often a user "bounces back" to the SERP immediately after selecting the document | • Individual user<br>• User group<br>• User population |
| Recommendation through links | (Weighted) number of links pointing to a certain document | • Links from all other pages<br>• Links from a group of pages, e.g., from topically relevant pages |
| Explicit ratings | Number of "likes," i.e., number of users who explicitly clicked on a "like" button such as Facebook's | • User group<br>• User population |

*Table 1. Implicit measures of credibility as applied by search engines*

## THE ROLE OF SEARCH ENGINE OPTIMIZATION

Search engine optimization (SEO) is the modification of Web page content, external links, and other factors, with the aim of influencing the ranking of the targeted documents. It must be stressed that with the well-known search engines, it is not possible to directly buy positions in the results lists, and advertising is separated from the organic results lists, and also more or less clearly labeled. However, with knowledge about search engines' ranking factors, it is possible to influence the results positions of the targeted documents.

While SEO can serve useful purposes, such as making documents findable in search engines through adding relevant keywords to the document, the techniques can also be used to manipulate search engines' rankings, and in the worst case, to help spam documents achieve top positions in the search results. So while search engine optimization does not directly affect the credibility of an individual document, in the SEO process, vast amounts of non-credible (i.e., pages of low quality) may be generated and be pushed into the search engines' results lists. The problem therefore does not lie in search engine optimizers producing non-credible documents, but often replacing more credible documents on the top of the results rankings.

Search engine optimization has now become an important business, as part of the online marketing industry. While traditionally, companies apply SEO techniques to boost their products and services in the search engines' rankings, SEO now goes further in that political parties, lobby organizations, and the like use these techniques, too. Even in the academic sector, there are efforts to optimize papers for academic search engines (so-called Academic SEO; see Beel, Gipp, & Wilde, 2010).

Looking at the credibility of search engines' results, SEO may not be problematic when products and services are concerned. One could argue that it does not matter whether a customer searching for some product will buy the best product at the best-possible price, or whether he buys an inferior product.

However, in the case of knowledge acquisition, the issue is different. Here, credibility matters, and SEO can severely influence the credibility of the results.

Search engine optimization tries to influence all implicit credibility judgments mentioned in Table 1, although selection behavior is an exception, at least to a certain degree, as it is hard to simulate real user behavior.

The real influence SEO has on areas where knowledge acquisition in concerned has still not been explored. However, there is some research on special sources that appear very often in search engines' results, such as Wikipedia articles (Lewandowski & Spree, 2011). From research on Wikipedia, we can see that lobbyists and public relations agencies try to influence public opinion by writing and editing articles. Furthermore, social media optimization aims at "optimizing" opinions expressed in social media. Again, there is a wide range of methods, ranging from Facebook popularity campaigns to writing fake reviews.

In summary, SEO and other "optimizing" strategies aim at influencing what users see in the top positions of search engine results pages. Agencies now see the control of the full first search engine results page as critical for success (Höchstötter & Lüderwald, 2011), whereas in the past, the aim was just occupying one top position.

We assume that the influence of search engine optimization will even rise in the coming years. However, it is difficult to project how search engines will react to increasing efforts to influence their results lists, when informational content is concerned. On the one hand, SEO can help to make good informational content visible in the search engines, and therefore, search engines such as Google even encourage content providers to use SEO techniques (Google, 2012). On the other hand, search engines need to keep the content displayed within the results lists credible, at least to a certain degree.

## UNIVERSAL SEARCH RESULTS

Universal Search is the composition of search engine results pages from multiple sources. While in traditional results presentation, results from just one database (the Web index) are presented in sequential order and the presentation of individual results does not differ considerably, in universal search, the presentation of results from the different collections (such as news, video, and images) is adjusted to the collections' individual properties. See figure 1 for an example, showing news and image results injected into the organic results list.

*Figure 1. Universal search results presentation in Google*

With universal search results, we see search engines turning away from pure algorithmic results. The idea of the general-purpose Web search engine was based on the facts that: (1) there are only low barriers for documents to be included in the index, and (2) due to the same algorithms applied to all documents in the index, every document has the same chance of being shown in a results list.

With universal search, however, these conditions do not necessarily hold true anymore. We see that search engines build highly-curated collections where only a very limited number of sources is included; thus, only search results from these sources are displayed. While a search engine's general Web index includes millions of sources (when considering every Web site to be a different source), the news index of a search engine only includes some hundred sources that a search engine considers as news media. While this surely does improve quality in terms of credibility (the documents themselves are considered of equal quality, and credibility judgments are made based on the source), low-quality documents may also be included in the universal search results due to questionable source selection (McGee, 2010).

Universal search surely improves search engines' results considerably. However, when showing results that are generated from special collections and are triggered by query words, there is also the danger of search engines favoring results from their own offerings or from partners (Edelman, 2010; Höchstötter & Lewandowski, 2009). However, there is still no consensus on whether search engines act as pure business entities, and therefore, there is no problem with such behavior (Granka, 2010) or whether search engines should be regarded as being responsible for providing "unbiased" results; however, that should be defined more precisely.

**CREDIBILITY IN SEARCH ENGINE EVALUATION**

Regarding the quality of search results, a vast body of research has been done over the years. In this section, we will discuss how information retrieval effectiveness tests relate to credibility of the search results. We will give a short overview of the testing methods, and then show how credibility assessments could be added to such tests.

Evaluation has always been an important aspect (and an important research area) of information retrieval (IR). Regarding Web search engines, established IR evaluation methods have been adapted to the context and been modified to suit Web searching. A good overview of newer approaches in Web search engine retrieval effectiveness evaluation is provided by Carterette, Kanoulas, & Yilmaz (2012).

In the retrieval effectiveness evaluation, two approaches need to be differentiated:

1. Retrieval effectiveness tests use a sample of queries and jurors to evaluate the quality of the individual results. These studies employ explicit relevance judgments made by the jurors.
2. Click-through studies analyze click data from actual search engine users. As users give their relevance judgments only through their selection behavior, we speak of implicit relevance judgments here.

Both approaches have merits. When using click-through data, researchers can rely on large quantities of data and can determine which results are preferred by the actual users of a search engine. The drawback, however, is that these decisions are based on the results descriptions on the SERPs that heavily influence users' results selections, and users choose only from some of the results presented. For example, a user would not read all the results descriptions and then choose a result from the third results page. On the contrary, he would rely on the first results presented by the search engine and choose from them.

The main advantage of classic retrieval effectiveness tests is that no data from the search engine providers are needed, and jurors can be explicitly asked for their opinions, so a researcher can go beyond decisions about whether an individual result is relevant or not. The drawback of such tests, however, is that such studies usually must rely on a relatively low number of queries and jurors, and results are seen as independent of one another. This can be illustrated by a user choosing a completely relevant result and who will therefore not need another relevant result that just repeats the information already given.

Regarding the judgment of the results, we speak of implicit relevance judgments in the case of click-through studies, and of explicit relevance judgments in the case of retrieval effectiveness tests. Relevance is a concept central to information science, but there is no agreed-upon definition of relevance in the field (Saracevic, 2007a, 2007b; Borlund, 2003; Mizzaro, 1997). However, this may not even be problematic when considering searches in curated collections and users being experts in their field; thus, the users can also consider credibility when assessing relevance. Tefko Saracevic, after many years of research on relevance, states: "Nobody has to explain to users of IR systems what relevance is, even if they struggle (sometimes in vain) to find relevant stuff. People understand relevance intuitively" (Saracevic, 2006, p. 9).

However, the question remains as to what extent this holds true when speaking of general-purpose Web search engine users. They surely see some results as relevant and others as not, but it is unclear to what extent they are able to discover errors or biases in the documents examined. Therefore, it is questionable whether judging the relevance of documents is truly sufficient when testing Web search engines.

Research examining the concepts underlying relevance found that many factors influence users' relevance judgments. Mizzaro (1997) reports research by Rees and Schultz, who found 40 variables influencing users' relevance decisions. Cuadra and Katter (1967) found 38 variables, and Barry and Schamber (1998) found 80 variables. Some studies (e.g., Chu, 2011) attempted to identify the most important variables in relevance judgments, but we still lack studies explicitly using individual variables, especially credibility, as criteria in search engine evaluation. Information science seems to regard credibility as just one part of the wider concept of relevance, which is central to the discipline.

One reason for using the more general concept of relevance instead of more specialized, credibility-oriented measures lies in the availability of jurors and the workload needed to judge the results' quality. When approaching relevance as stated in the quote by Saracevic above, relevance judgments are relatively easy to obtain, and decisions can be made relatively quickly.

Most search engine studies use students as jurors, and sometimes faculty also help to judge the documents (Lewandowski, 2008, p. 918-919). While this is convenient, it is questionable whether these jurors are indeed able to evaluate the credibility of the results, as they may lack deeper understanding of the topics to be researched and may invest too little time in checking for credibility. A further concern is that search engine evaluations increasingly "crowd-source" relevance judgments, i.e., the tasks are distributed over the Internet to a large group of paid-for jurors. While this allows for larger studies, the jurors often lack commitment to the tasks and are interested in completing them in the least amount of time possible.

When assessing credibility of the results, it would be best to use expert jurors and to ask them to examine the results thoroughly. However, expert jurors are difficult to contact and must also be paid for their services. Therefore, studies using such experts are, apart from some very topically specialized studies, quite rare.

As can be seen, the main problem with using credibility in search engine evaluations is recruiting suitable expert jurors. It may seem simple for experts to judge documents, but often, a rather large amount of time is needed for researching every statement from a document. Even when the criteria for judging credibility are unambiguous, the time-consuming process of examining tens or even hundreds of pages (as done in retrieval effectiveness tests) is a major barrier not only for adding credibility to search engine tests, but also to adding other factors.

Keeping this in mind, we suggest taking a four-part approach for incorporating credibility judgments into search engine retrieval effectiveness tests:

1. Use actual search engine users to judge results relevance (as done in retrieval effectiveness studies).
2. Let the same users judge the credibility of the results. Add questions on the type of credibility the jurors assign to the individual result: presumed credibility, reputed credibility, surface credibility, and experienced credibility.
3. Use automatic approaches to judge credibility. These approaches can be used to at least give indications on reputed quality and surface quality. However, it should be noted that search engine rankings themselves could be seen as judgments of reputed quality.
4. Use expert judged to judges to judge the credibility of individual results.

This multi-dimensional approach can be used to comparing laypersons' and experts' rating of results credibility. For search engine vendors, information on whether the results presented are seen as credible by both groups is of value for improving their ranking algorithms. For internet researchers, such information would be valuable for determining whether search engines are suitable tools for research in certain areas, e.g. health topics.

In contrast to studies on the credibility of search engines results, there is a vast body of research on frameworks for judging the credibility of Web content (Rieh, 2002; Wathen & Burkell, 2002; Fogg, 2003; Hilligoss & Rieh, 2008; for an overview, see Rieh & Danielson, 2007). A challenge for research on Web search engines lies in combining the advantages of established retrieval effectiveness test methods with judgments on credibility. One can argue that asking jurors only to judge relevance is realistic, in that users generally do not think much about the credibility of the documents, but instead rely on the intermediary (the search engine) and its ranking, but to understand search engines as dominant tools in knowledge acquisition, further research on their overall quality is needed (Lewandowski & Höchstötter, 2008). Research from other disciplines, such as journalism or lexicography, where frameworks for measuring credibility exist can help to build better credibility measurements for Web search engines (Lewandowski & Spree, 2011).

## THE RESPONSIBILITY OF WEB SEARCH ENGINES FOR CREDIBLE INFORMATION

It could not be stressed enough how search engines have become major tools for knowledge acquisition. This can be seen when looking at the number of queries entered into the general-purpose search engines every day. According to research from ComScore (ComScore, 2010), more than 131 billion searches were conducted in December 2009 alone. The question that arises from these figures is what role search engines should play in promoting credible Web documents. While their algorithms surely honor credibility to a certain degree (see above), credibility is still not explicitly a concept applied in search engines. Vertical search engines, i.e., search engines that do not provide a "complete" index of the Web but instead focus on a specific area (such as news, jobs, or scholarly articles) are one step toward source credibility, as the sources are—in many cases—handpicked.

However, this vertical search approach is not applicable to general Web searches, as the Web is simply too large to base search engines on human-curated collections. However, the combination of the general-purpose Web index with vertical indices is a way to achieve higher credibility of the overall results. Problematic with this approach, however, is that neither inclusion criteria for vertical indices nor lists of sources included are currently published by the search engines. Furthermore, such vertical indices exist only for some areas, and these are not necessarily the areas where curated indices are deemed most useful (e.g., health, law, and authoritative information from governments).

The question remains as to whether search engines should be seen as pure business entities that can decide whatever content they want to present and in which way, or whether they have a special responsibility to provide users with credible results that are free from bias.

## FUTURE RESEARCH DIRECTIONS

From the argumentation above, we derive three major areas for further research about credibility in Web search engines:

1. Applying credibility in search engine evaluations

   Search engine evaluation traditionally focuses on relevance, which incorporates a multitude of concepts. While relevance is a suitable concept for judging the overall quality of a result to a query or information need and can easily be used in evaluations using laypersons, it is hard to differentiate between the factors that make a result relevant. Furthermore, jurors often cannot judge on the credibility of the results, and may even not notice obvious factual errors.

   Therefore, search engine evaluations should focus more on the credibility of search engine results. While this surely makes evaluations more complicated, as results must be checked for errors and bias, the results would help a great deal in finding what quality of results users are able to view when using search engines.

2. Increasing users' awareness of credibility issues in search engines

   It is well known that users do not invest much cognitive effort in the formulation of queries and the examination of search engine results (Höchstötter & Koch, 2009; Machill, Neuberger, Schweiger, & Wirth, 2004). Furthermore, search engines are often seen as responsible for the content of the results provided, and search engines such as Google are referred to as the source for information, where in reality, they are simply intermediaries between the searcher and the content providers. The result from this is a great trust in search engines, especially in Google, as the most popular search engine.

   Further research should focus on how users' information literacy can be improved, especially concerning search engine-based research, where credibility is a major concept to be explored. Furthermore, users' awareness of how search engines rank the results and how search engine optimization can influence (the credibility of) the results should be raised.

3. Building collections of credible sources

   In vertical search engines, collections of credible sources are already built. These collections can be grouped into automatically generated collections (such as Google Scholar's collection of scholarly articles), and handpicked collections (such as Google News). Especially for critical domains such as health and law, search engines' results could significantly benefit from the latter. Microsoft's Bing search engine shows how health-related information from credible sources can be incorporated into search engines' results. However, the sources deemed credible need to be disclosed in such collections. While Bing uses a limited set of health authorities, in other cases such as news, it remains unclear how the collection is built and what the criteria for inclusion are.

4. Search engine vendors' responsibility for the quality of the results
   It remains unclear how much of a problem (lack of) credibility is in Web search engines. From examples from such diverse areas as credit-related information, health, law and politics, we know that search engines at least sometimes present biased and/or non-credible information. Research examining to what extent search engines provide users with such content, and how users react to such results, is needed.

## CONCLUSION

Credibility is an important concept relevant to search engine evaluation and for search engine providers as well. However, as we have seen in this chapter, in neither area is credibility explicitly used. Search engine providers could benefit from building more human-curated collections of credible sources, as they would help to lead users to the best results available on the Web. Another way of promoting credibility in search engine results pages is to give the users better support for judging credibility. Kammerer and Gerjets (2012) suggest the following three areas: (1) Reducing the prominence of search results ranking, (2) increasing the prominence of quality-related cues on search engine results pages, and (3) automatic classification of search results according to genre categories. While (1) is mainly applied in "alternative" search engines experimenting with results presentation and visualization, we can see that the major Web search engines increasingly give quality-related cues on the SERPs through, e.g., author and freshness information. There are some experiments regarding automatic genre classification, but this is not applied in the major Web search engines, yet.

Search engine evaluation should also focus more on credibility, instead of relying on jurors who are themselves not experts in a field and work under time restrictions. As search engines are employed by many millions of users every day for a multitude of purposes, and these users trust in the results provided by the engines, more effort should be put into researching not only the technical means of search engines, but also their impact on knowledge acquisition.

**ADDITIONAL READING SECTION**

## KEY TERMS & DEFINITIONS

Index
The database on which a search engine is based. A search engine can have multiple indices. The core index is the index of Web documents ("Web index"), where only low inclusion barriers apply and as many documents as possible (and economically maintainable) are included. This index can be accompanied by other (vertical) indices, which are based on a selection of sources.

Ranking signal
Search engines use many signals that together form the ranking algorithm. A signal refers to a single criterion that can be used in document ranking. Based upon definition, search engines apply some hundreds or even thousands of signals in their rankings.

Search engine optimization (SEO)
Search engine optimization (SEO) is the modification of Web page content, external links and other factors, aiming to influence the ranking of the targeted documents.

Search engine results page (SERP)
A search engine results page is a complete presentation of search engine results; that is, it presents a certain number of results (determined by the search engine). To obtain more results, a user must select the "further results" button, which leads to another SERP.

Universal Search
Universal Search is the composition of search engine results pages from multiple sources. While in traditional results presentation, results from just one database (the Web index) are presented in sequential order and the presentation of individual results does not differ considerably, in universal search, the presentation of results from the different collections (such as news, video, and images) is adjusted to the collections' individual properties.

Vertical search engine
Contrary to the general-purpose search engine, a vertical search engine focuses on a special topic. General-purpose search engines often enhance their results with results from vertical search indices, such as news, video, or images.